# Estimating the evolution and the content fractions of baryonic gas for Luminous Infrared Galaxies

M. N. Al Najm, Ahmed H. Abdullah and Y. E. Rashed ⋆

*Department of Astronomy and Space, College of Science, University of Baghdad, Baghdad, Iraq*

**ABSTRACT**
Luminous Infrared Galaxies (LIRGs) play a crucial role in understanding of galaxy evolution. The present study examined 82 LIRGs, using data taken from the Sloan Digital Sky Survey (SDSS), NASA/IPAC Extragalactic Database (NED), and HyperLEDA to explore their gas fractions and optical properties. The analysis of data highlights the relationship between molecular-to-atomic mass of hydrogen gas ratio $M_{H_2}/M_{H\,\textsc{i}}$ and morphological types, gas mass fractions, and galaxy characteristics such as color and luminosity. The results showed that the regressions between $M_{\rm dust} - M^{*,V}$ and $M_{\rm dust} -$ SFR are not quite flat (when correlation coefficient $> 0.5$), which indicates a decrease in the dust-to-stellar content ratio as the gas is consumed and transformed into stars, and also a relatively flat trend for $M_{\rm dust} - M^{*,V}$ and $f_{\rm dust,bar} - M^{*,V}$. Moreover, as the star's mass declines, the total gas mass fraction ($f_{\rm gas}$) increases quickly, with a high negative correlation coefficient of $-0.7$ and a regression of $-0.85$. Therefore, it can be inferred that galaxies with a high gas fraction ($f_{\rm gas}$) are either accreting gas at a rate sufficient to meet their energy requirements for star formation or converting gas into stars less effectively. According to the findings, the gas exhaustion time in these galaxies quickly reduces as the stellar mass increases, with a significant negative correlation coefficient of $-0.7$ and a regression that is a nearly linear regression of $-0.9$. On the other hand, when the baryonic gas mass fraction grows, which makes up the majority of the baryonic gas, grows, the gas depletion time increases quickly.

## 1 INTRODUCTION

Luminous infrared galaxies (LIRGs) have been a focal point of galaxy research since their discovery over three decades ago. They have a high infrared luminosity ranging from $10^{11}$ to $10^{12}$ solar luminosities within the wavelength range of 8 $\mu$m to 1000 $\mu$m (Pérez-Torres et al. 2021). It is obvious that luminous infrared galaxies derive energy from star formation and an active galactic nucleus (AGN). Massive stars and/or AGNs radiate ultraviolet photons and convert a significant amount of dust and gas in the galaxy into infrared radiation. Many studies suggest that mergers may serve as a primary mechanism driving star formation and possibly fueling the AGN (Madau & Dickinson 2014; Jiang et al. 2024). Mergers of gas-rich galaxies are important for understand the relationship between galaxies and their central supermassive black holes (SMBHs) (Zhuang & Ho 2023). Understanding the evolution of LIRGs is crucial for explaining the existence of numerous luminous quasars at high redshift, as merger processes were believed to be more common in the early universe (Sanders & Mirabel 1996; Romano et al. 2021).

A newly identified class of galaxies, discovered by the Infrared Astronomy Satellite (IRAS) survey, emits most of its luminosity in the infrared (IR). The IRAS produced the original IRAS point source catalog by surveying 96 per cent of the sky, constrained by a completeness threshold of approximately 0.5 Jy at 12, 25, and 60 $\mu$m and approximately 1.5 Jy at 100 $\mu$m Duaa & Al Najm (2022); Shaker & Al Najm (2024).

The formation of any galaxy and its subsequent evolution is entirely dictated by its gas supply (Al-baqir & Ahmed 2024). In ellipticals, the gas supply might be turned into stars right after formation, while in tidally induced starbursts, it remains dispersed until an event that enhances the rate of cloud collisions occurs (e.g. LIRGs). The gas mass fraction $f_{\rm gas}$, defined as a ratio of the mass of the gas to the total mass of the galaxy, is the most important quantity that determines the evolutionary state of the galaxy. Chemical and photometric evolution are directly proportional to the rate at which the gas is consumed in a galaxy. Bell & de Jong (2000) showed that the metallicity of spiral galaxies aligns with a straightforward closed-box chemical evolution model, and this model also predicts the colors of the primordial stellar population (Schombert, McGaugh & Eder 2001).

Recent studies and surveys using advanced infrared instruments have provided clear evidence of the ongoing increase in the star content of galaxies during the cosmic era (Fontana et al. 2006). Based on the studies by Giallongo et al. (1998) and Papovich, Dickinson & Ferguson (2001), who established methods for determining the stellar masses in galaxies at high redshifts, numerous surveys have highlighted that overall the composition of stars within the cosmos as suggested by the mean density of stellar masses, has increased over cosmic time.

Stars are thought to be born in cold molecular hydrogen ($H_2$) gas clumps that have undergone gravitational collapse, which can later divide into gaseous cores capable of forming individual stars. Cooling of the interstellar medium (ISM) is considered a key process in conventional star formation models, crucial for the formation of $H_2$ gas clumps or cores, and intimately linked with star formation efficiency. Although current knowledge about star formation in galaxies has advanced considerably, the details of this process have not been fully elucidated yet (Fontana et al. 2006; Abdullah & Kroupa 2021). The interstellar gas found within spiral galaxies may constitute a significant proportion of their baryonic mass. Many studies indicate a strong correlation between the combined mass of stars and gases and the kinematic measurements of the total mass content, such as the widths of H I-line profiles (Kourkchi et al. 2022; Park et al. 2022). Utilizing the amount of dust present to extract insights about the gas content is feasible when the ratio of dust to gas is known. The majority of investigations targeting the assessment of gas content in distant galaxies at high redshifts have depended on observations of carbon monoxide (e.g. Genzel et al. 2010; Tacconi et al. 2013).

⋆ E-mail: yassir.e@sc.uobaghdad.edu.iq

These studies have made possible of the investigation the relationships between the mass of molecular gas and star formation rate (SFR) across different periods in the cosmos. Nevertheless, these observations require a significant amount of time and are influenced by uncertain factors, such as the CO-to-$H_2$ conversion factor, which remain poorly defined, especially for starburst or metal-poor galaxies (Santini et al. 2014). Bellocchi et al. (2022) conducted a detailed comparative analysis of the molecular, stellar, and ionized gas distributions and their sizes. Additionally, they juxtaposed local galaxies and distant high-redshift systems with low-redshift luminous infrared galaxies for a comprehensive analysis to provide a broader view of the comparison samples. The present study demonstrates that galaxies at higher redshifts possess larger proportions of gas compared to low-redshift luminous infrared galaxies. Consequently, star formation may be triggered through a series of interactions and mergers take place within extended disks or filaments display a sufficiently low molecular gas surface density. These processes involve physical mechanisms akin to those observed in the central kiloparsec of LIRGs. (Stanley et al. 2023) conducted a survey of CO(J = 1 - 0) for distant galaxies between redshifts 2 to 4 that emit intense infrared radiation due to an active star formation within dusty environments. As it turns out, thermal gas pressure is directly involved in the star formation process of these galaxies. They investigated the possible connection between thermal gas pressure and star formation efficiency and also discovered that thermal gas pressure directly affects star formation in these galaxies due to strong evidence of the correlation.

## 2 SAMPLES AND DATA SET

A set of 82 luminous infrared galaxies were selected from the study (Shaker & Al Najm 2024). The data were chosen to provide as large a database as possible to investigate the relationship between the gas fraction and optical properties of luminous infrared galaxies in the visual band. In this study, the data were collected from three photometric extragalactic surveys: Sloan Digital Sky Survey (SDSS) sky server Data Release 17 (DR17), NASA/IPAC Extragalactic Database (NED) surveys and HyperLEDA database. The data used in our study was sourced from established survey databases, including SDSS, NED, and HyperLeda, where essential preprocessing steps, such as spectral extraction and calibration, had already been conducted. The SDSS sky server DR17 provides color filters at green g-band ($\lambda$ = 4770 Å), red r-band ($\lambda$ = 6231 Å), and extinction in the r-band magnitude (Er). The following data are mapped from the HyperLEDA database: the morphological type, numerical code for the revised (de Vaucouleurs) morphological type T for each source, isophotal level 25-mag/arcsec$^2$ in the B-band diameter of galaxies ($logd_{25}$) expressed in arcmin, apparent total blue-band, V-band magnitudes reduced to the Third Reference Catalogue of Bright Galaxies (RC3) system B and V, effective $(B - V)$ color which is given essentially in the UBV system, the inclination to the line-of-sight $i$ [in degrees] which is determined from the apparent flattening and the morphological code type, H I magnitude line $m_{21c}$, which is corrected for self-absorption effect, $V_{3k}$ the radial velocity relative to the Cosmic Microwave Background (CMB) radiation, and line width at 50 per cent of the maximum $W_{50,HI}$ in km s$^{-1}$ measured from the radio line width. From NED surveys, the redshift z of the galaxies was collected. This set formed the basic sample of the present study. The first ten datasets in this study are listed in Table 1 and all of these parameters are of luminous infrared galaxies. Fig. 1 is a sample of LIRG object in which *NGC 7679* was observed in the radio band using the ALMA interferometric radio telescopes.[1]

## 3 DETERMINING PARAMETERS

The fundamental parameters of the study star masses $M_{star}$, star formation rate (SFR), dust, baryonic, and dynamic masses were discussed. Furthermore, an overview of the basic formulas for baryonic gas fractions and some methods for calculating them was provided. The total mass gas $M_{gas}$, stellar masses, star formation rate, dust masses, total mass gas, gas metallicities, and baryonic gas mass fraction are utilized in the following ways:

(1) By applying the post-Hubble relation, the photometric physical distance $D$ (in Mpc) of the (LIRGs) sources is calculated (Karachentsev et al. 2006):

$$D = \frac{V_{3k}\left(\frac{1-(q_0-1)V_{3k}}{2c}\right)}{H_0} \quad (1)$$

Using the Sun's motion coefficients with the cosmic microwave background (CMB), the radial velocity $V_{3k}$ in the CMB regime is calculated from the heliocentric radial speed $V_h = cz$, where $z$ represent the redshift referred to in Section 2 and $c$ is the speed of light. The cosmological constant ($\Omega_m = 0.308$, $\Omega_\Lambda = 0.692$), and the value of the deceleration parameter ($q_0 = -0.55$) are consistent with the typical cosmological model that includes cold dark matter as a component. The Hubble constant $H_0$ was taken to be 67.8 km s$^{-1}$ Mpc$^{-1}$ (Abd Al-Lateef & Mahdi 2022; Zamal & Al Najm 2023).

(2) For luminous infrared galaxies, the total integrated H I flux intensity, and $F_{HI-21}$ were obtained from the 21-cm line in absorption and radiation and can be determined by neutral hydrogen (H I) disc system mass, $M_{HI}$ (Guo et al. 2023; Chandola et al. 2024):

$$M_{HI}(M_\odot) = 2.36 \times 10^5 \times D^2 \times 10^{-0.4(m_{21c}-17.4)} \quad (2)$$

where $m_{21c}$ is the corrected apparent magnitude in the H I line at $\lambda$ = 21 cm, according to the formula (Paturel et al. 1997):

$$m_{21c} = m_{21} - 2.5 \log\left(\frac{k/\cos(i)}{1-\exp(k/\cos(i))}\right) \quad (3)$$

where, $k$ is a free parameter equal to 0.031 (Paturel et al. 1997).

(3) A greater far-infrared (FIR) luminosity can be seen in illuminated infrared objects. By taking 100 μm flux $F_{100}$ and a dust

---
[1] More details can be found on the ALMA Science Portal: https://almascience.nrao.edu/aq

**Table 1.** The major parameters of luminous infrared galaxies.

| Name | T | log$d_{25}$ (arcmin) | g | r | $E_r$ | $B - V$ corrected | i (deg.) | $V_{3k}$ (km s$^{-1}$) | $m_{21c}$ | $W_{50,HI}$ (km s$^{-1}$) | z |
|---|---|---|---|---|---|---|---|---|---|---|---|
| NGC 23 | 1.2 | 1.19 | 12.817 | 12.031 | 0.09 | 0.88 | 40.5 | 4222 | 15.41 | 383 | 0.01523 |
| NGC 5257 | 3.2 | 1.17 | 14.915 | 14.966 | 0.06 | 0.53 | 62.1 | 7089 | 15.09 | 464.5 | 0.02267 |
| NGC 5258 | 3.2 | 1.17 | 14.099 | 13.372 | 0.06 | 0.58 | 34.2 | 7083 | 14.66 | 437.5 | 0.01523 |
| NGC 5936 | 3.2 | 1.07 | 13.048 | 12.285 | 0.09 | 0.60 | 19.3 | 4179 | 15.63 | 180 | 0.0133 |
| NGC 7469 | 1.1 | 1.14 | 13.080 | 12.384 | 0.16 | 0.49 | 30.2 | 4545 | 16.58 | 216 | 0.01627 |
| NGC 7591 | 3.6 | 1.21 | 13.635 | 12.663 | 0.24 | 0.93 | 66.9 | 4585 | 14.13 | 383 | 0.01654 |
| NGC 7679 | −1.3 | 1.09 | 13.081 | 12.600 | 0.15 | 0.50 | 59.3 | 4777 | 15.45 | 274 | 0.01715 |
| NGC 7769 | 3.0 | 1.26 | 12.868 | 12.075 | 0.17 | 0.80 | 73.2 | 3866 | 15.60 | 303 | 0.01405 |
| NGC 7817 | 4.1 | 1.52 | 20.678 | 19.487 | 0.13 | 0.88 | 90.0 | 1959 | 13.93 | 403.6 | 0.01532 |
| UGC 03973 | 3.0 | 1.13 | 14.466 | 15.475 | 0.16 | 0.50 | 36.7 | 6765 | 16.15 | 172 | 0.02221 |

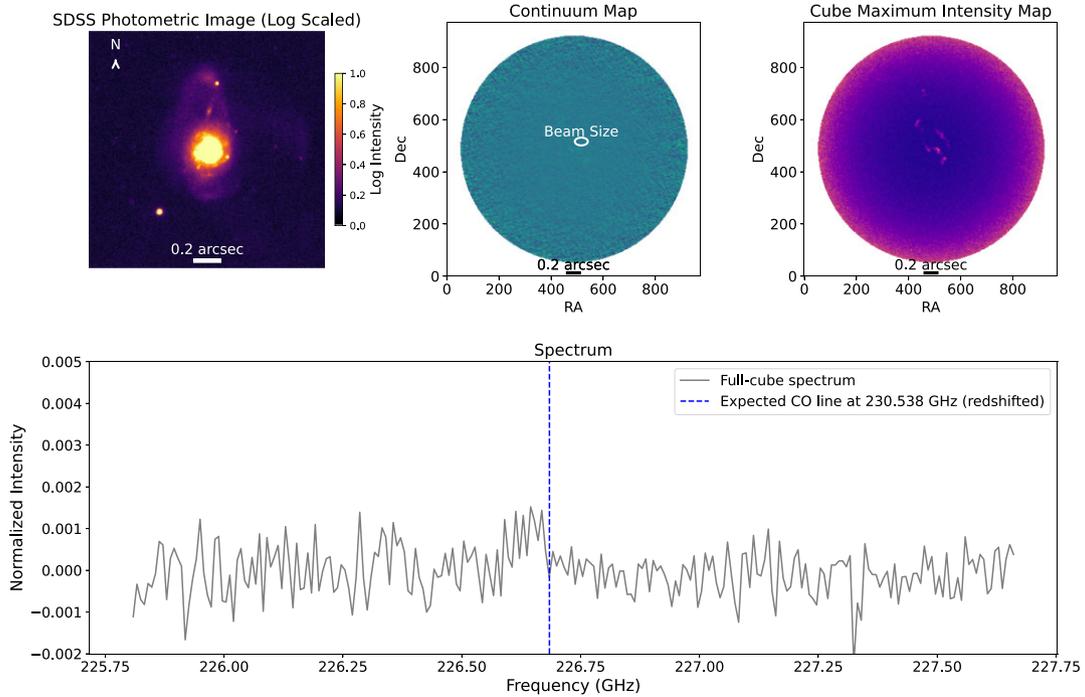

**Figure 1.** ALMA radio data for the galaxy NGC 7679, showing the continuum map (top left), SDSS photometric image taken at g-band (top center), cube maximum intensity map (top right), and a representative spectrum extracted from the center pixel (bottom). The continuum map reveals the overall distribution of radio emission, while the cube maximum map highlights the peak intensity across the spectral cube. The spectrum displays the frequency range covered by this dataset, though the CO(j = 2-1) line at 230 GHz is notably absent, suggesting a lack of detectable molecular gas at this transition.

temperature $T_{\rm dust}$, the dust mass can be calculated as follows (Combes et al. 2011):

$$M_{\rm dust} = 4.8 \times 10^{-11} \frac{F_{\nu_o} D^2}{(1+z) k_{\nu_r} B_{\nu_r}(T_{\rm dust})} \quad (4)$$

where $B_{\nu_r}$ is the Planck radiation law at the rest frequency $\nu_{\rm rest} = \nu_o(1+z)$, $F_{\nu_o}$ is the calculated FIR flux taken in Jy, and $k_{\nu_r}$ is a mass opacity parameter of 25 cm$^2$ g$^{-1}$ at rest-frame 100 µm, assuming a frequency dependence of $\beta = 1.5$ (Combes et al. 2011). It should be noted that $F_{60}$ µm and $F_{100}$ µm fluxes are used to determine the dust temperature (Lavezzi et al. 1999).

$$T_{\rm dust} = 49 \left( \frac{F_{60\,\mu m}}{F_{100\,\mu m}} \right)^{0.4} \quad (5)$$

The mass of dust $M_{\rm dust}$ is estimated as follows (Combes et al. 2011; Duaa & Al Najm 2022):

$$M_{\rm dust}(M_\odot) = 5(1+z)^{-(4+\beta)} F_{100\,\mu m} D^2 \left[ \exp\left( \frac{144(1+z)}{T_{\rm dust}} \right) - 1 \right] \quad (6)$$

or in the form:

$$M_{\rm dust}(M_\odot) = 5(1+z)^{-(4+\beta)} F_{100\mu m} D^2 \\ \times \left[ \exp\left( 2.94 \left( \frac{F_{100}}{F_{60}} \right)^{0.4} (1+z) \right) - 1 \right] \quad (7)$$

(4) The total contained mass (or dynamical mass $M_{\rm dyn}$) within a radius $R_{25}$ of a spherically symmetric structure can be expressed in solar units as follows (Saintonge et al. 2011):

$$M_{\rm dyn} = \frac{V_c^2 R_{25}}{G} \quad (8)$$

Hence, using the H I linewidth as a reference, $V_c$ is the circular velocity, given by $V_c = \frac{W_{50,HI}}{2\sin i}$, the speed widths of the profile (in km s$^{-1}$) at 50 per cent peak maximum are specified by $W_{50,HI}$, and $i$ is the inclination of the galaxy system disk (where $i$ is the angle of the galaxy's polar axis to the line of sight, as measured from HyperLeda photometry) (Paturel et al. 2003). $G$ denotes the universal constant of gravity, and given by $G = 4.3 \times 10^{-6}$ kpc km$^2$ s$^{-2}$ M$_\odot^{-1}$. The galaxy system disk radius, or $R_{25}$ in kpc, is the value of a surface brightness equivalent to 25-B mag arcsec$^{-2}$.

(5) Stellar mass is an essential physical parameter to be evaluated. Since the stellar mass-to-light ratio, $\Upsilon^*$, is highly dependent on the metallicity and star formation activity of a galaxy, one cannot simply estimate the stellar mass of a galaxy from its visual luminosity. The optical colours were broadly employed to calculate $M^*/L_\lambda$ ratios (de los Reyes et al. 2015; Al Najm, Rashed & AL-Dahlaki 2024; Jasim & Al Najm 2024). Stellar mass is estimated from $M^* = \Upsilon^* \times L_\lambda$ (at the wavelength $\lambda$ specified by the applied filter) for the visual (V-band) magnitude of luminous infrared galaxies with the mass-to-light ratio $\Upsilon^*$ assessed using synthesis models with $(B - V)$ colors (McGaugh & Schombert 2015; McGaugh, Schombert & Lelli 2017): For the V-band ($\lambda = 5500$ Å):

$$\Upsilon^{*,V} = 10^{-0.628+1.305(B-V)} \quad \text{in units} \quad M_\odot/L_\odot \quad (9)$$

In general, the emitted luminosity $L_\lambda$ is computed from V-band data (Johnston 2011; Schneider 2015; Al Najm et al. 2024):

$$L_{(V-band)} = 10^{-0.4(M_V - M_{V,\odot})} L_{V,\odot} \quad (10)$$

where $M_{V,\odot} = 4.82$ mag. is the absolute magnitude of the Sun in the visual band.

The absolute magnitude is represented by the symbol the symbol $M_X$, where $X$ is the filter that has been considered. According to the definition, $M_X$ is the same as the apparent magnitude ($m_X$) of a source that is 10 parsecs away from us. As opposed to apparent magnitude, a source's absolute magnitude is hence independent of its distance. The relationship between apparent and absolute magnitude is determined (Schneider 2015).

$$M_X = m_X - 5\log\left(\frac{D_{pc}}{10\,pc}\right) \quad (11)$$

or

$$M_X = m_X - 5\log(D_{Mpc}) - 25 \quad (12)$$

So, the absolute magnitude $M_V$ for the V-band is calculated as:

$$M_V = m_V - 5\log D_{Mpc} - 25 \quad (13)$$

where $m_V$ is the apparent magnitude of galaxies in the V-band. Thus, the stellar mass in units of $M_\odot$ is given by:

$$M^{*,V} = \Upsilon^{*,V} \times 10^{-0.4(M_V - M_{V,\odot})} \quad (14)$$

(6) The total gas mass $M_{gas}$ includes the combined value of the helium mass $M_{He}$, the molecular gas $H_2$, and the atomic mass $M_{HI}$ of neutral hydrogen HI, and referred to as the total gas content $M_{gas}$ (Goddy et al. 2023; Al Najm et al. 2024; Ravi, Douglass & Demina 2024):

$$M_{gas}(M_\odot) = M_{H_I} + M_{H_2} + M_{He} = 2.36 \times 10^5 \times D^2 \times 10^{-0.4(m_{21c}-17.4)}$$
$$+ 0.07 M^{*,V} + 0.33(M_{H_I} + M_{H_2}) \quad (15)$$

By considering a mass fraction of 25 per cent Ravi et al. (2024), it can roughly estimate the mass of helium ($M_{He}$):

$$M_{He}(M_\odot) = \frac{0.25}{1 - 0.25}(M_{H_I} + M_{H_2}) = 0.33(M_{H_I} + M_{H_2}) \quad (16)$$

This is the estimated amount of helium in the intergalactic medium, which is consistent with the predictions of Big Bang nucleosynthesis (Cooke & Fumagalli 2018).

One frequent molecular tracer used in astrophysics to investigate gaseous dynamics, star formation (SF), and the physical conditions of galaxies is carbon monoxide (CO). Each rotational transition in the CO emissions lines (represented by the letter J) is a quantum 'jump' between rotational energy states. CO should be emitted at $\nu_{rest} \sim$ 115.271 GHz ($\lambda \sim 2.60$ mm), $\nu_{rest} \sim 230.538$ GHz ($\lambda \sim 1.30$ mm), and $\nu_{rest} \sim 345.796$ GHz ($\lambda \sim 0.87$ mm) according to the predicted emissions (CO $J = 1 \to 0$, CO $J = 2 \to 1$, and CO $J = 3 \to 2$) He et al. (2020). Due to its stability and symmetry, molecular gas is challenging to be determined directly. Instead, it is often tracked using CO emissions, which are subsequently converted utilizing a scaling relationship between CO flux and molecular gas mass. However, there is much uncertainty in this relationship (Goddy et al. 2023). The sample of luminous infrared galaxies (LIRGs) typically exhibits weak or undetectable CO line emission, and possible fluctuations in the CO–H$_2$ conversion factor complicate its interpretation. It was detected a rare rotational transition in the CO-line emission such as $^{12}$CO$(J = 1 \to 0)$, $^{12}$CO$(J = 2 \to 1)$, $^{12}$CO$(J = 3 \to 2)$ in our dataset as shown in the radio spectrum of NGC 7679 Fig. 1. By combining two scaling relationships–one that converts the star mass to the star formation rate and another that converts the star formation rate to the molecular gas mass. McGaugh, Lelli & Schombert (2020) estimate the molecular gas content ($M_{H_2}$) and determine that the molecular gas mass in a galaxy is well approximated by '7 per cent' of the stellar mass, meaning that $M_{H_2} = 0.07 M^*$ (Goddy et al. 2023). As a result, this simplifying approximation was used to give the relationship between the molecular gas mass and the stellar mass in the V-band as follows:

$$\log M_{H_2}(M_\odot) = \log M^{*,V} - 1.16 \quad (17)$$

(7) The Star formation rate (SFR): The amount of gas that was transformed into stars throughout the galaxy's history of star construction (Rashed et al. 2013; Abdullah et al. 2019; Rashed 2023). The relationship between SFR and $M_{star}$, also known as the main sequence of star-forming luminous infrared galaxies or the star-forming sequence. The star formation rate component is integral to the determination of stellar mass (McGaugh et al. 2017; Lisenfeld et al. 2023):

$$M_{star} = \int_{\tau_f}^{\tau_u} SFR\, dt = \langle SFR \rangle \tau_G \quad (18)$$

where $\tau_u$ is the age of the universe in years, $\tau_f$ is the galaxy's development time, and $\tau_G = \tau_u - \tau_f$ is the galaxy's lifetime. The mean amount of star formation is $\langle SFR \rangle$, is given by $M_{star}/\tau_G$.

Assuming $\tau_G = 13$ Gyr, the association between SFR and star mass for V-band aligned with the line of constant star formation McGaugh et al. (2017):

$$\log SFR = \log M^{*,V} - 10.11 \quad (19)$$

Here, the SFR is in units of $M_\odot$ yr$^{-1}$ and $M^{*,V}$ is in units of $M_\odot$.

(8) The amount of time needed to deplete the observed supply of cold hydrogen gas at the current star formation rate can be calculated as follows (McGaugh et al. 2017):

$$\tau_{dep} \text{ (in unit yrs.)} = \frac{M_{gas}}{SFR} \quad (20)$$

(9) The Hubble morphological classifications affect molecular gas fractions relative to atomic gas masses (Young & Knezek 1989;

McGaugh & de Blok 1997):

$$\frac{M_{H_2}}{M_{H_I}} = 3.7 - 0.8T + 0.043T^2 \quad (21)$$

where $T$ represents internal numerical codes. The collection of morphological descriptions is the origin of this parameter.

(10) Baryonic mass Tully–Fisher relations was established in the V-band, where the baryonic mass mass $M_{bar}$ is the sum of stellar and interstellar medium gas masses (McGaugh 2005; McGaugh et al. 2020; Goddy et al. 2023):

$$M_{bar} = M^{*,V} + M_{gas} \quad (22)$$

This mass produces the Baryonic Tully- Fisher Relation (BTFR), where the mass directly includes all observed baryonic components: stars and gas. This mass related agreeably with a galaxy's rotation speed. By substituting equation (14) into equation (22), the yield will be the following formula:

$$M_{bar} = \Upsilon^{*,V} \times 10^{-0.4(M_V - M_{V,\odot})} + (M_{H_I} + M_{H_2} + M_{He}) \quad (23)$$

(11) Gas mass fractions are produced of three primary mechanisms govern the gas composition of galaxies: gas accumulation through input from the intergalactic medium, depletion of gas resulting from star formation, and outpourings because of active galactic nuclei and the creation of stars. Since the gas fraction denotes these opposing influences, its development of $f_{gas}$ over time contains crucial information about their relative strength (McGaugh & de Blok 1997; Bothwell et al. 2013; Roman-Oliveira, Rizzo & Fraternali 2024). The baryonic gas mass fraction is

$$f_{gas} = \frac{M_{gas}}{M_{bar}} = \frac{M_{gas}}{M_{gas} + M^{*,V}} \quad (24)$$

(12) The entire mass of baryons in a galaxy divided by the galaxy's total mass is known as the baryon/dynamical mass fraction. The overall galaxy mass can be determined using either the total dynamical mass (referred to as the disc baryon fraction) or the total halo mass (referred to as the halo baryon fraction) (Bradford, Geha & Blanton 2015). The fraction of baryons that have formed in a galactic disk is determined by the disc baryon fraction, also known as the concentrated baryon fraction of galaxies. For both scenarios, the galaxy's baryonic/dynamics fraction ($f_{bar,dyn}$) has been established by the formula:

$$f_{bar,dyn} = \frac{M_{bar}}{M_{dyn} + M_{bar}} = \frac{M^{*,V} + M_{gas}}{M_{dyn} + M^{*,V} + M_{gas}} \quad (25)$$

## 4 RESULTS AND DISCUSSION

### 4.1 $M_{H_2}/M_{H_I}$ ratio as a function of morphological Hubble code type

The molecular-to-atomic mass ratio of hydrogen gas ($M_{H_2}/M_{H_I}$) and the morphological Hubble code type (T) is shown in Fig. 2. According to the Second Reference Catalogue of Bright Galaxies or RC2, the revised morphological type's numerical code (de Vaucouleurs) is specified. The Hubble type is referred as follows: E = -5, $S0^-$ = -3, $S0^0$ = -2, $S0^+$=-1, S0a=0, Sa=1, Sab=2, Sb=3, Sbc=4, Sc=5, Scd=6, Sd=7, Sdm=8, Sm=9, and Ir = 10. In addition to the difficulties with morphological categorization, the scatter may result from galaxies of varied brightness, spiral arm intensity, and surroundings within each type. In this study, it was compelled to fit the late types galaxies to obtain $(M_{H_2}/M_{H_I}) \approx 0$ as T approaches 10 (refer to equation (21)). The significant quantity of the true dispersion in each category is more concerned. According to equation (21), the statistical measure of the ratio of molecular gas content, is not entirely correlated with T. For Sa-Sb, the average value of the molecular-to-atomic gas ratio is steadily dropped by a factor of ∼ 2 as a function of morphological type (Fig. 2). Despite the significant variation in the molecular-to-atomic gas ratio within a certain type, a definite pattern shows that the $M_{H_2}/M_{H_I}$ varies with the Hubble type. The scatter could be caused by difficulties in the morphological categorisation, in addition to including galaxies of different brightness, environments, and spiral arm amplitudes within each type. It was noted that late-type galaxies (such as Sd-Sm-Irr) have molecular mass five times less than atomic gas, whereas early-type galaxies (such as E-S0) have molecular mass five times more than atomic mass gas. Depending merely on the H I abundant, the late types are assumed to be gas-abundant. If massive amounts of molecular gas is lost in late forms, it will only construct new gas-rich galaxies. There is no basis to assume that the overall ratio of molecular to atomic gas is constant as a function of morphological Hubble type. This is because the abundances of atomic and molecular gas within galaxies is changed greatly. Therefore, the observed variation in $(M_{H_2}/M_{H_I})$ with type is considered as an inherent characteristic. As a result, even though the statistical details may be enhanced in the CO data, the overall conclusions are quite reliable The findings are highly compatible with the studies conducted McGaugh & de Blok (1997) and Young & Knezek (1989).

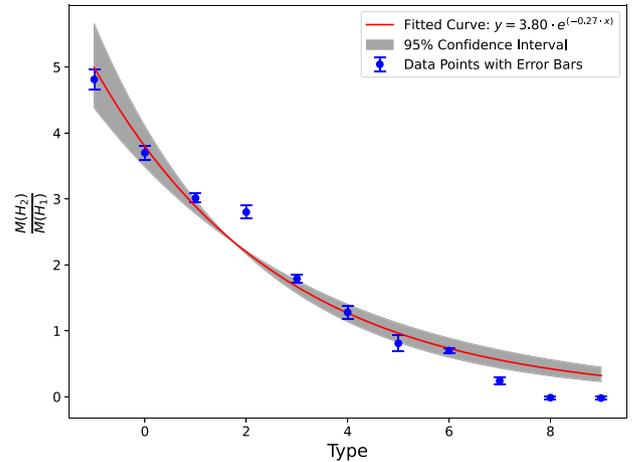

**Figure 2.** The molecular to atomic hydrogen gas mass ratio as a function of Hubble code type. As is typical, the real dispersion at an identified Hubble type is underestimated by the mean error. The line shows our fit to the database.

### 4.2 Gas mass fraction as a function of colour

The integrated $(g-r)_{corrected}$ colour corrected for Galactic extinction was utilized in the SDSS DR17, and the $(g-r)_{corrected}$ was calculated using the relationship: $(g-r)_{corrected} = (g-r)_{observed} - E_r$. The variations in integrated $(g-r)_{corrected}$ galaxy colours as a function of the baryonic gas mass fraction and baryonic/dynamic fraction (Fig. 3a). For luminous infrared galaxies, it was noted that there is no clear correlation between these variables under study. For Fig. 3(b), the results indicated a negative connection between the gas mass fraction and the colour index $(B-V)_{corrected}$, with a regression −0.3 and a standard error of 0.1. The low SFR might cause this in these galaxies; the combined stellar and gaseous surface density is below the minimum requirement for star formation, even though they include massive quantities of H I. The gas fractions are plotted versus the absolute magnitude $M_B$ of the luminous infrared galaxies in blue-band characteristics as shown in Fig. 3(c). The link between gas fraction and $M_B$ is significant with a partial association coefficient of 0.51 and a regression with a standard error of $0.620 \pm 0.11$. The relationship between baryonic/dynamics fraction and $M_B$ is weakly negative, with a partial coefficient correlation of $-0.3$ (Fig. 3d). The gas fraction, $M_B$, and $(B − V)_{\rm corrected}$ are related to each other.

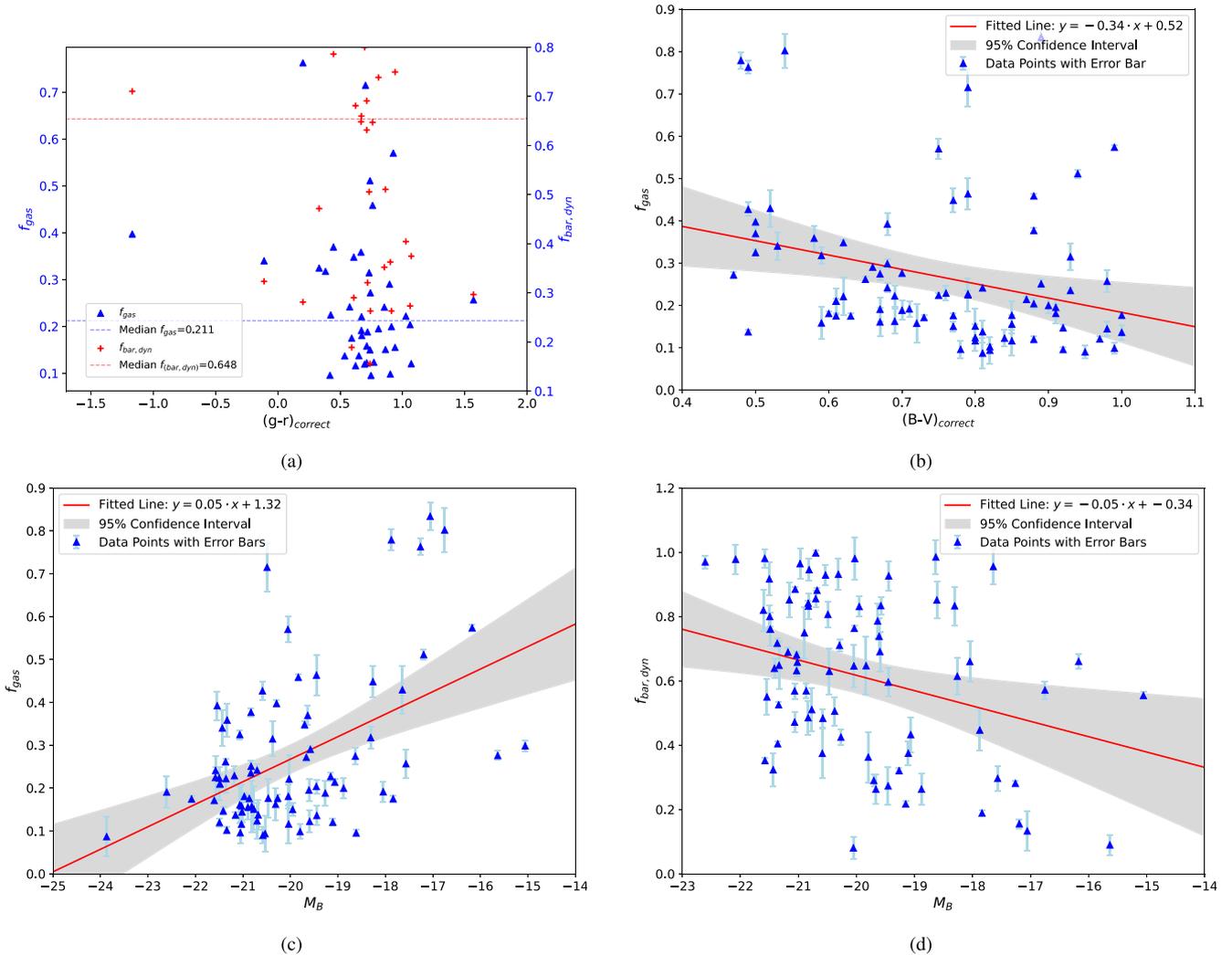

**Figure 3.** (a) The baryonic gas mass fraction and baryonic/dynamic fraction as a function of $(g − r)_{\rm corrected}$ colour and (b) The baryonic gas mass fraction as a function of $(B − V)_{\rm corrected}$ colour. (c) The baryonic gas mass fraction plotted against absolute magnitude in the blue band and (d) The baryonic/dynamic fraction mass fraction plotted against $M_B$.

### 4.3 Stellar mass versus $M_{\rm dust}$ and SFR

It is essential to test the relationship between the stellar mass and dust mass because these galaxies contain an abundance of cosmic dust, with an average value of dust mass of approximately $M_{\rm dust} \approx 5.4 \times 10^6$ M$_\odot$. The dust mass fraction of the sample can also be computed as: $f_{\rm dust,bar} = \frac{M_{\rm dust}}{M_{\rm bar}} = \frac{M_{\rm dust}}{M_{\rm gas}+M^*}$. Fig. 4(a) displays the fit findings and the dust mass as a function of the star mass at various redshifts ($0.001 \leq z \leq 0.051$). The correlations of log $M_{\rm dust}$ against log $M^{*,V}$ and log $M_{\rm dust}$ against log SFR were linearly fitted. The findings indicate that the relationships between $M_{\rm dust} - M^{*,V}$ and $M_{\rm dust}$ − SFR are not nearly flat, with a correlation coefficient and regression of $> 0.5$ (a regression of 0.5 or less indicates a flat relationship). This relationship which considered by other investigators (Santini et al. 2014) becomes flatter and occasionally vanished, suggesting at least partly caused indirectly by other factors. Particularly, the relationship between $M_{\rm dust}$ and SFR, along with the link between $f_{\rm dust,bar} - M^{*,V}$ contribute to the $M_{\rm dust} - M^{*,V}$ correlation (refer to Fig. 4(b) and Fig. 4d). The results of this investigation are compatible with theoretical predictions concerning the development of the dust mass, which predict a pretty flat trend for $M_{\rm dust} - M^{*,V}$ and $f_{\rm dust,bar} - M^{*,V}$, or a decrease in the dust-to-stellar content ratio as the gas is consumed and converted into stars. On the other hand, this can be noticed in high stellar mass for luminous infrared galaxies with an average value ($M^{*,V} = 3.5 \times 10^{10}$ M$_\odot$). Additionally, this result might suggest that a significant amount of dust is lost in these systems that exhibit an insignificant dust content with a mean value of $M_{\rm dust} \approx 5.4 \times 10^6$ M$_\odot$. The results clearly show that the average dust mass proportion to the baryonic gas/dust mass $f_{\rm dust-bar}$ (which includes both gases and stars), is low, at about 0.00018. They supports earlier findings that dust is being converted into stars in these galaxies. Furthermore, a strong relation is existed between SFR and mass gas $M_{\rm gas}$ with a linear regression of 0.7 as shown in Fig. 4(c).

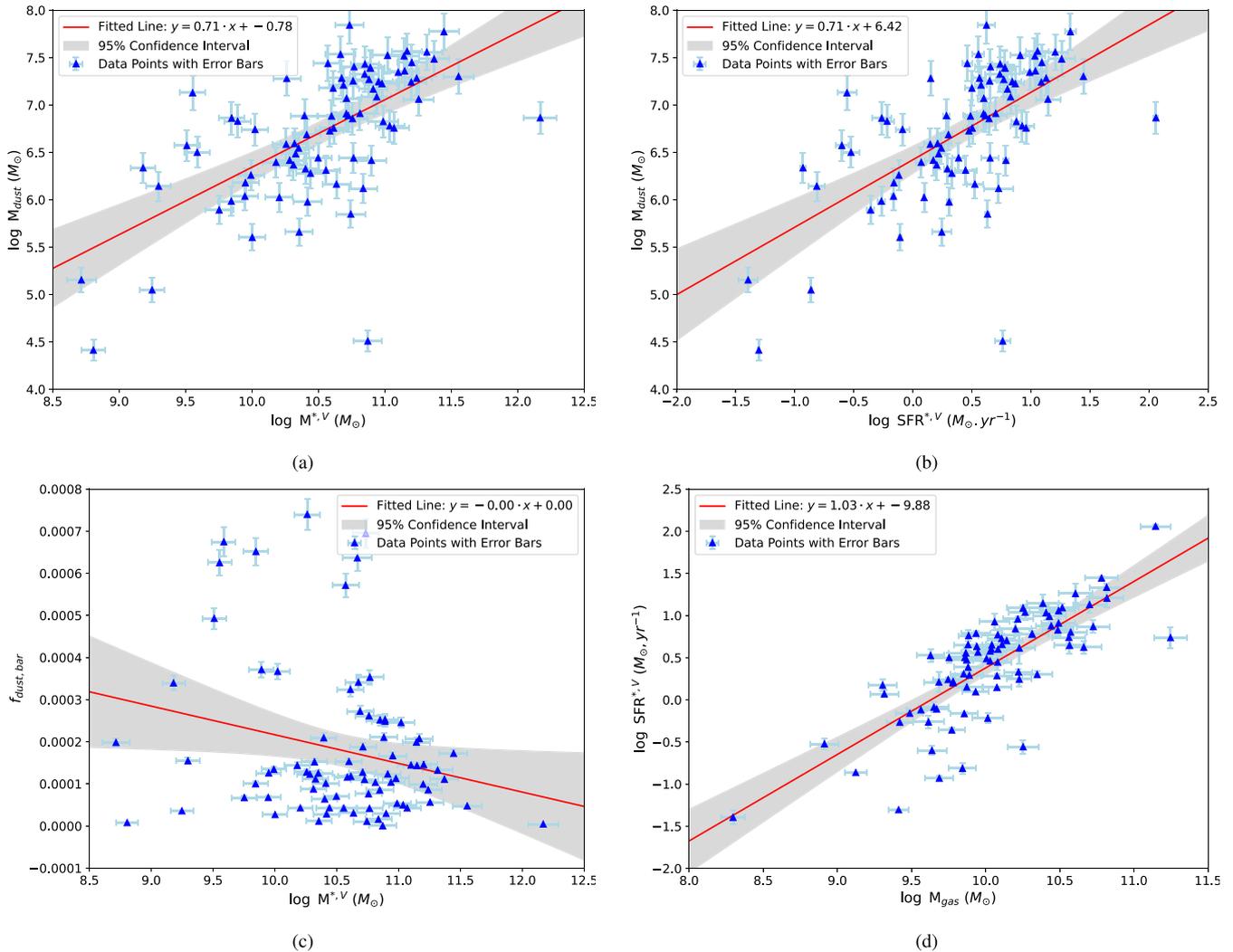

**Figure 4.** Dust mass plotted against (a) stellar mass, and (b) star formation rate. (c) show the baryonic gas/dust mass as a function of stellar mass and (3d) display the relation between star formation rate and total gas mass at different redshifts ($0.001 \leq z \leq 0.051$).

### 4.4 Gas fraction and baryonic/dynamic fraction versus stellar mass

Fig. 5(a) a illustrates the relationship between stellar mass and gas mass fraction. The total gas mass fraction ($f_{\rm gas}$) has a strong correlation coefficient of $-0.7$ and a regression of $-0.85$, increases rapidly as the star's mass decreases. With $M_{\rm gas} < M^{*,V}$, the luminous infrared galaxies have comparatively low total gas masses. They are lower than the incremental fit to the LIRGs database extension. In fact, at the scale $M^{*,V} = 3.5 \times 10^{10}\,{\rm M_\odot}$, there seems to be an evolution in the behavior. For the scale $M_{\rm gas} \approx 10^{10}\,{\rm M_\odot}$, LIRGs create a series of gas-poor galaxies below this mass. The pattern $M^{*,V} \approx 3.5\,M_{\rm gas}$ does not deviate much from a fixed value in an initial calculation. It was noticed in the system of bright infrared galaxies, the ratio of fraction gas mass to stellar mass reduces at a gradient of approximately $M^{*,V} \geq 10^{10}\,{\rm M_\odot}$ with $f_{\rm gas} \leq 0.3$. This maybe because luminous infrared galaxies are close to depleting their gas reservoir for star formation and have low total gas mass fractions. It is easy to envision that galaxies gain a maximum ($M_*$) along an evolutionary trend of decreasing proportion of gas mass as their gas supply runs out. The evolution's typical timeframe may be a function of mass, with $M^{*,V} \approx 10^{10}\,{\rm M_\odot}$ serving as the present shutdown mass. A series of vibrant, luminous infrared galaxies that contain enough gas to continue generating stars for a very long time lie above this mass. However, beyond that, there are more aged varieties of these galaxies, most of whose star-forming ability has passed. Conversely, the mass of the stars and the baryonic/dynamic fraction positively are correlated. Despite the increase in the star mass, the overall baryonic/dynamic fraction rises (as shown in Fig. 5b). The relationship seen in the right-hand panel of Fig. 5(a) between dynamical mass, total baryonic content (the sum of gas and star masses), and stellar mass behaves somewhat differently from the total gas mass fraction that was just reviewed. A baryonic-dynamic fraction ($f_{\rm bar/dyn}$), with a regression of 0.34 and a weak partial correlation coefficient of 0.3, is a significantly flatter function of stellar mass than the total gas mass fraction ($f_{\rm gas}$). It exhibits only a slight regular increase with $M^{*,V}$. Considering the almost linear correlation between $f_{\rm gas}$ and $M^{*,V}$, $f_{\rm bar/dyn}$ should normally have a flat correlation with $M^{*,V}$, with a range value of 0.08 to 0.99 for the whole mass of the star. In Fig. 5(c), the scattering of the hydrogen atomic mass-to-luminosity ratio ($M_{HI}/L_V$) is obvious. The mean value of the samples $M_{HI}/L_V = 0.74$ which spans the range ($0.005 \leq M_{HI}/L_V \leq 13$) indicating that $M_{HI}$ depends on the Hubble morphological category. The ratio of the gas mass to luminosity is a commonly used metric to evaluate a galaxy's gas richness. For luminous infrared galaxies, the relationship of $M_{HI}/L_V$ versus stellar mass is shown in Fig. 5(c). $M_{HI}/L_V$ and $M^{*,V}$ have a substantial association, with a regression of $-0.7$ and a correlation coefficient of $-0.62$. The gas generation and its evolutionary condition can be well-indicated by this robust $M_{HI}/L_V - M^{*,V}$ relationship. By measuring the average value of the ratio of the mass of natural hydrogen to the luminosity $M_{HI}/L_V$, which is roughly 0.7, one may determine that the gas in these galaxies was depleted and converted into stars. The quantity of gases in these galaxies is indicated by the mass of neutral hydrogen, and the equations (14) and (19) suggest that luminosity is an accurate measure of star formation. Stars comprise most of the baryonic material in the sample of luminous infrared galaxies. The phrase 'gas-rich galaxies' thus describes this reversal, from gas-mass supremacy for these galaxies to stellar-mass predominance in disks of luminous infrared galaxies. Thus, it may be concluded that galaxies with a large gas fraction ($f_{gas}$) are either accreting enough gas to replenish their star-making energy needs or these galaxies are less efficient when turning their gas into stars.

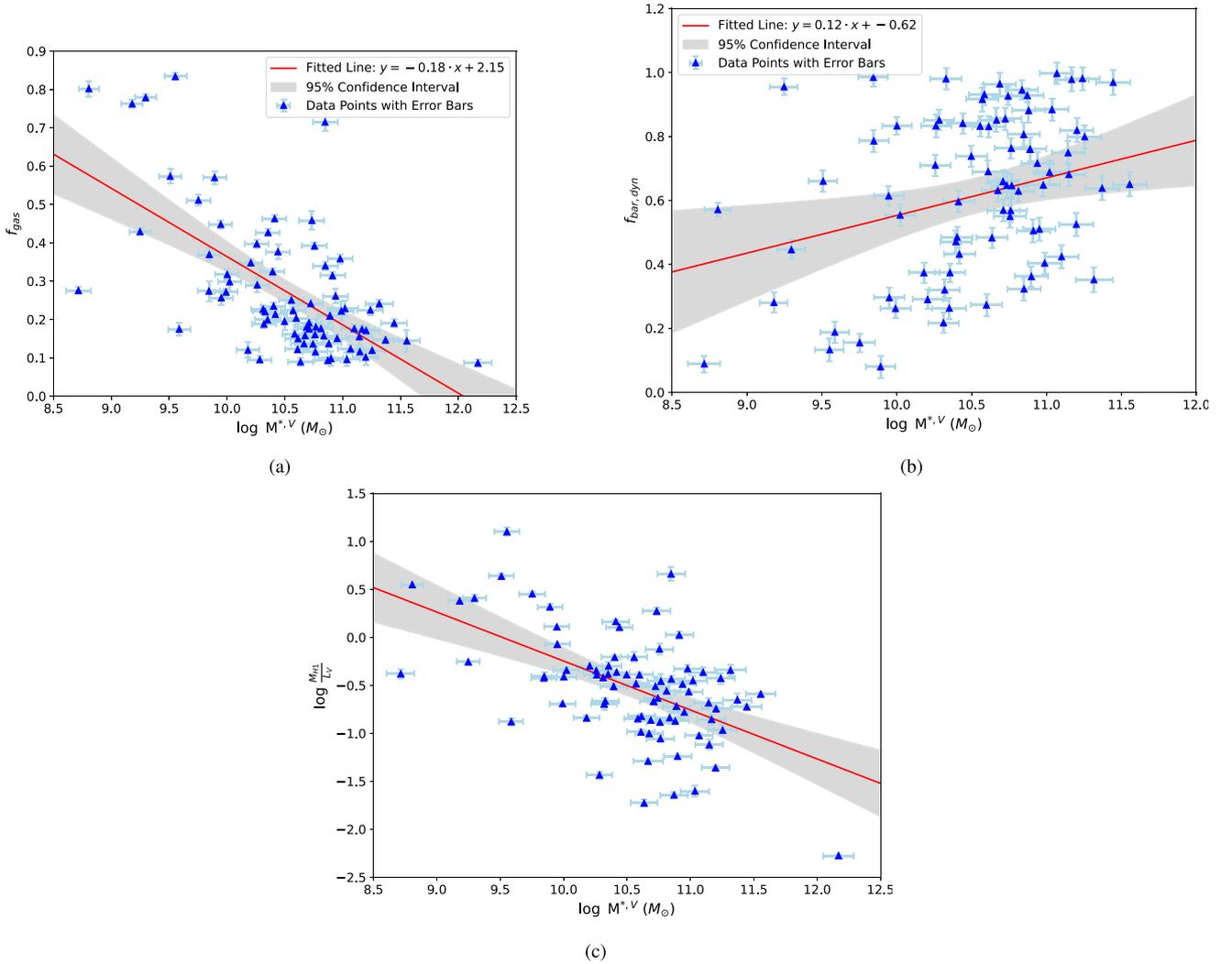

**Figure 5.** (a) Represent the gas Mass fraction vs. stellar mass and (b) Baryonic/Dynamic mass fraction vs. stellar mass in solar mass, while (c) Represent the relationship between $M_H/L_V$ and stellar mass in solar mass.

### 4.5 Relationship between gas fraction and gaseous depletion periods $\tau_{dep}$

Three basic mechanisms govern the gas composition of galaxies: gas accretion through input from the interstellar medium, usage of gas due to star formation, and jets resulting from active galactic nuclei and star formation. The gas fraction represents these opposing factors, and as such, the evolution of $f_{gas}$ encodes essential information about the relative importance of these factors across time. The formula $\tau_{dep} = \frac{M_{gas}}{SFR}$ determines how long it takes to exhaust the observable gas supply at the present rate of star formation. As a function of gaseous depletion time with baryonic gas mass fraction and star mass ($\tau_{dep} - f_{gas}$ and $\tau_{dep} - M^{*,V}$), this is displayed in Figs 6(a) and (b). Luminous infrared galaxies are sometimes thought to require outside accumulation due to a short depletion time, Fig. 6(a) show that these galaxies are distinct. The results show a gas exhaustion time in these galaxies that decreases rapidly with increasing stellar mass, with a regression close to linear at $-0.9$ and a strong negative correlation coefficient of $-0.7$. Conversely,

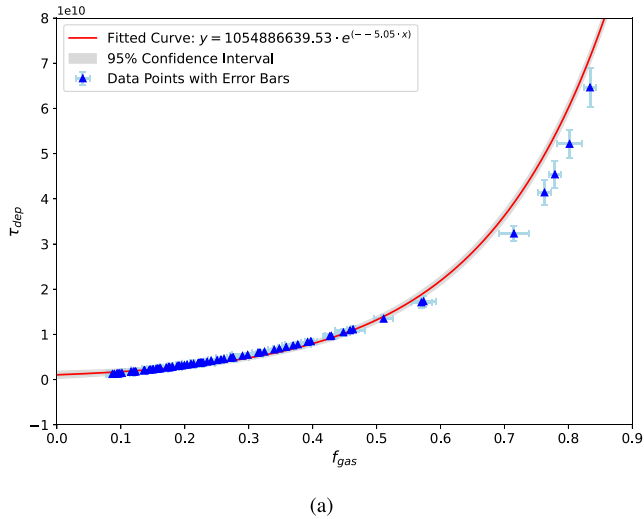 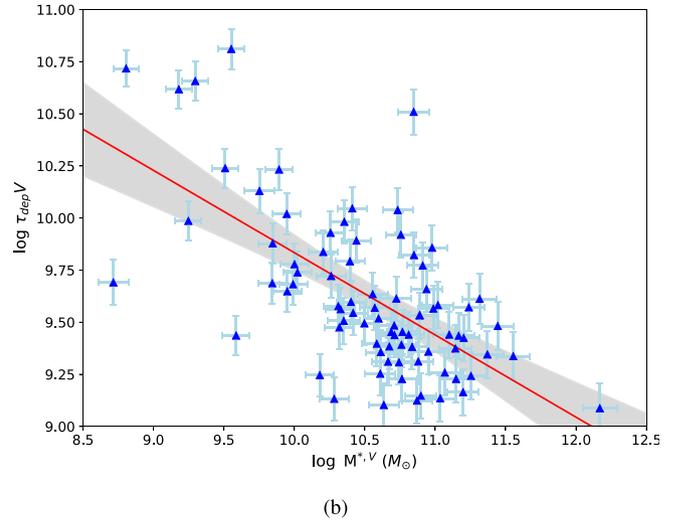

(a)          (b)

**Figure 6.** Gas depletion time in years plotted against and (a) baryonic gas mass fraction and (b) stellar mass in solar mass.

Fig. 6(b) depicts a rapid increase in the gas depletion time as the fraction of baryonic gas, which constitutes increasing in most of the baryonic gas. This increase follows a linear regression equal to one and exhibits a highly robust positive partial correlation coefficient approaching 0.9. Figs 6(a) and (b) illustrates that the largest galaxies ($M^{*,V} \geq 10^{10} M_\odot$ and $f_{gas} > 0.5$) have gas exhaustion times smaller than the Hubble time, with the most massive galaxies finishing star formation first. However, growing galaxies with less mass ($M^{*,V} < 10^{10} M_\odot$ and $f_{gas} > 0.5$) may not need to form new stars for up to hundreds of giga years ($\tau_{dep} \sim 100$ Gyr). They can generate stars at the measured rates for a very long time, and their gas depletion durations are considerably beyond a Hubble epoch. According to the findings, the shorter exhaustion durations of luminous galaxies could mean they have evolved closer to completion, indicating that a larger proportion of their initial gas has formed stars. This aligns with the comparison to turning off the main chain; star production is slowing down in huge galaxies as the gas supply gets closer to running out. Short depletion durations do not support the argument of the theory that bright infrared galaxies remain significant amounts of new accumulated gas. It's more likely that giant galaxies with short depletion times indicate an ongoing process rather than living in a unique era where LIRGs are only now depleting their gas supplies. Massive galaxies, such as giant galaxies, have the strongest actual star formation rates among regular galaxies in the nearby universe and are still actively producing stars. Their unique star formation rates, nevertheless, seem to be decreased.

The fragile relationship between the gas exhaustion time and the baryonic/dynamic mass fraction with a regression of $-0.27$ is due to the presence of the total dynamic mass being included in the baryonic mass fraction, and this indicates the dependence of the gas exhaustion time with the total gas mass (which includes neutral and molecular gases and the element helium), and with the rate of stellar formation, which in turn is related to the stellar mass, as shown in equation 20.

## 5 GOODNESS OF THE FIT

The accuracy of the results which represents by the coefficient of determination ($R^2$) and statistical significance (p-value) is shown in Table 2. The correlation is usually regarded as statistically significant

**Table 2.** Summary of $R^2$ and p-values for different figures and variables.

| Figure No. | Variables (Dependent–Independent) | $R^2$ | p-value |
|---|---|---|---|
| Fig. 3 a | $f_{gas} - (g-r)_{corrected}$ | 0.093 | 0.14 |
| Fig. 3 a | $f_{bar,dyn} - (g-r)_{corrected}$ | $\sim 0.04$ | $\sim 0.7$ |
| Fig. 3 b | $f_{gas} - (B-V)_{correct}$ | 0.11 | 0.007 |
| Fig. 3 c | $f_{gas} - M_B$ | $\sim 0.3$ | $\sim 10^{-5}$ |
| Fig. 3 d | $f_{bar,dyn} - M_B$ | 0.17 | $\sim 0.007$ |
| Fig. 4 a | $\log M_{dust} - \log M^{*,V}$ | 0.43 | $\sim 10^{-5}$ |
| Fig. 4 b | $\log M_{dust} - \log SFR$ | 0.43 | $\sim 10^{-5}$ |
| Fig. 4 c | $f_{dust,bar} - \log M^{*,V}$ | $\sim 0.2$ | $< 0.00047$ |
| Fig. 4 d | $\log SFR - \log M_{gas}$ | 0.43 | $< 10^{-7}$ |
| Fig. 5 a | $f_{gas} - \log M^{*,V}$ | $\sim 0.5$ | $< 10^{-7}$ |
| Fig. 5 b | $f_{bar,dyn} - \log M^{*,V}$ | $\sim 0.17$ | 0.009 |
| Fig. 5 c | $\log \frac{M_{HI}}{L_V} - \log M^{*,V}$ | 0.33 | $\sim 10^{-7}$ |

if the p-value is less than a predetermined significance level ($< 0.05$).

## 6 DEPENDENCE OF METALLICITY ON THE EVOLUTION OF BARYONIC GAS FRACTION

As metallicity increases, the chemical evolution of the baryonic gas fraction is expected to decrease according to the following equation (McGaugh et al. 2020).

$$\phi = 1 - (y + Z) \quad (26)$$

The initial helium abundance ($y_p$) at the boundary of zero oxygen abundance can be estimated by a linear regression of the mass fraction of helium ($y$) as a function of oxygen abundance concerning hydrogen (O/H). The helium mass fraction ($y$) is an indicator of hydrogen gas (O/H) abundance of oxygen (McGaugh et al. 2020):

$$y = y_p + \frac{dy}{d(\text{O/H})}(\text{O/H}) \quad (27)$$

So

$$\phi = 1 - \left(y_p + \frac{dy}{d(\text{O/H})}(\text{O/H}) + Z\right) \quad (28)$$

According to Fukugita and Masahiro's Fukugita & Kawasaki (2006) estimation, $y_p$ increases from $0.234 \pm 0.004$ to $0.250 \pm 0.004$.

They also revealed that the initial phase of stellar absorption significantly affects $dy/dZ$, with the initial value of $dy/dZ = 4 - 5$ decreasing to $1 \pm 1$. This lower value agrees with the fraction suggested by the typical solar model. In this study $dy/dZ = 1.1$ and $y_p = 0.25$ were employed. When metallicity is represented by oxygen $(dy/d(\text{O/H}) = 18.2(dy/dZ)$ (Izotov & Thuan 2004)), the result is:

$$\frac{dy}{d(\text{O/H})} = 18.2 + \frac{dy}{dZ} \qquad (29)$$

This yields the equation:

$$\phi = 0.75 - 38.2(\text{O/H}) \qquad (24)$$

When combined with the mass-metallicity correlation found in de los Reyes et al. (2015), we obtain:

$$\phi = 0.75 - 38.2 \left( \frac{M^{*,\text{V}}}{M_o} \right)^\alpha \qquad (30)$$

where $\alpha = 0.22$ and $M_o = 1.5 \times 10^{24} M_\odot$ (Chae et al. 2020; McGaugh et al. 2020).

The factor $\eta = \phi^{-1}$ defines the intergalactic hydrogen fraction and gas components other than H I. Additionally, gas phases often contribute very little, typically in the visual disc where all pertinent values are acquired. Rather than the commonly considered solar value ($\eta = 1.4$), an initial hydrogen proportion ($\eta = 1.33$) has been investigated in numerous studies Schombert et al. (2001); Lelli et al. (2019); Al Najm, AL-Dahlaki & Alkotbe (2023); Di Teodoro et al. (2023); Al Najm et al. (2024). In the baryonic Tully–Fisher relationship inquiry, the flat transformation method $M_{\text{gas}} = (1.33 - 1.4) M_{\text{H I}}$ is utilized to combine the proportions of metals and molecular gas to account for additional components to the gas mass Ball et al. (2023). Based on the mineral mass link provided in the formula (25), Table 3 displays the analysis data for the mass dependence of the gas hydrogen fraction (18). In the current study, the values of $\eta$ obtained from several trackers were contrasted with the mass of the stellar component $M^{*,V}$ computed previously using the mass-to-light ratio equation (14). The statistical results show that the minimum and maximum values of the factor $\eta$ are ranged between $\eta_{\min} \approx 1.36$ and $\eta_{\max} \approx 1.5$, with the mean value of $\eta \approx 1.4 \pm 0.026$. The variance of the star's mass is large, about 2.5, while the variance of the hydrogen gas fraction component is minimal, about 0.03. The results in this study are significant, due to the reinforcement of the hypothesis that can adopt the mass dependence of the gas fraction ($\eta$) at a value range from 1.33 to 1.4.

## 7 CONCLUSIONS

Through the samples of 82 LIRGs that were focused on in the study, the conclusion of this study as follows:

(i) The correlation between the molecular-to-atomic mass ratios of hydrogen gas ($M_{H_2}/M_{H I}$) and the morphological Hubble type points out systematic variation in gas content via various galaxy kinds. The early-type galaxies (E and S0) showed an ($M_{H_2}/M_{H I}$) ratio of approximately five times higher than the late-type (Sc, Sd, Sm, and Irr).

(ii) The mean value of the ($M_{H_2}/M_{H I}$) ratio is approximately $\leq 0.4$ for Sd-Sm-Irr galaxies and $\sim 5$ for E-S0 galaxies. It was showed that the ratio of molecular to atomic gas among luminous infrared galaxies declines by a factor of $\sim 2$ as a function of morphological Hubble type.

(iii) No clear correlation was found between integrated $(g - r)_{\text{corrected}}$ colors and the variables studied. Moreover, no significant correlation was found between gas fraction and $(g - r)_{\text{corrected}}$ color index, likely because of the low SFR. A clear relationship was found between gas fraction and absolute magnitude $M_B$ in the blue band, with a partial association coefficient of 0.51 and a regression of $0.620 \pm 0.11$. The correlation between baryonic/dynamic fraction and $M_B$ was weakly negative. These results confirm the interrelation of gas fraction, $M_B$, and $(g - r)_{\text{corrected}}$ in LIRGs.

(iv) The relationships between $\log M_{\text{dust}}$ vs. $\log M^{*,V}$, and $\log M_{\text{dust}}$ vs. $\log \text{SFR}$, present a clear correlation with coefficients and regressions of $> 0.5$, indicating non-flat relationships. These findings, agree with theoretical predictions, propose that dust-to-stellar content ratios decline as gas is converted into stars. High stellar mass LIRGs ($M^{*,V} \approx 3.5 \times 10^{10} M_\odot$) exhibit low dust content, supporting the concept of dust conversion into the stars. The dust mass average was proportion to the baryonic mass about 0.00018. Furthermore, a strong relationship was found between SFR and $M_{\text{gas}}$, with a linear regression of 0.7.

(v) A strong correlation was found between stellar mass and a gas mass fraction ($f_{\text{gas}}$), with a coefficient correlation of $-0.7$ and a regression of $-0.85$, indicating stellar mass increases, while $f_{\text{gas}}$ decreases. LIRGs overall have low total gas masses, mostly at stellar masses around $M^{*,V} \approx 3.5 \times 10^{10} M_\odot$. For $M_{\text{gas}} \approx 10^{10} M_\odot$, LIRGs shows low gas. As stellar mass increases to $M^{*,V} \geq 10^{10} M_\odot$, $f_{\text{gas}}$ drops below 0.3, suggesting these galaxies are depleting their gas reservoirs. There is a positive correlation between stellar mass and baryonic/dynamic fraction ($f_{\text{bar/dyn}}$) with a regression of 0.34 and a correlation coefficient of 0.3. The correlates between ($M_{H I}/L_V$) and stellar mass, have a regression of $-0.7$ and a correlation coefficient of $-0.62$, indicating ongoing gas depletion and star formation. These findings confirm that LIRGs with high gas fractions are either replenishing their gas or are less efficient in converting gas to stars.

(vi) The results showed that the gas exhaustion time in (LIRGs) decreases rapidly with increasing stellar mass (regression $\approx -0.9$, correlation coefficient $= -0.7$), but increases with higher baryonic gas fractions (regression $\approx 1$, correlation coefficient $= 0.9$). The Massive galaxies ($M^{*,V} \geq 10^{10} M_\odot$, $f_{\text{gas}} < 0.5$) have gas exhaustion times shorter than the Hubble time, suggesting they are closer to completing star formation, while Smaller galaxies ($M^{*,V} < 10^{10} M_\odot$, $f_{\text{gas}} > 0.5$) can afford star formation for much longer periods ($\tau_{\text{dep}} \sim 100$ Gyr). These findings suggest that massive LIRGs have evolved significantly, converting most of their gas into stars, while less massive galaxies continue to form stars. The weak relationship between gas exhaustion time and baryonic/dynamic mass fraction (regression $= -0.27$) indicates that gas exhaustion time depends on the total gas mass and star formation rate, which is related to stellar mass.

(vii) The study final analysis showed that the value of the factor $\eta$, which determines the mass fraction of gas and hydrogen components in the interstellar medium, has a mean of $\eta \sim 1.4 \pm 0.026$ and ranges from $\eta_{\min} \sim 1.36$ to $\eta_{\max} \sim 1.5$.

## ACKNOWLEDGEMENTS

The authors wish to express their gratitude for the resources provided by the HyperLeda database (http://leda.univ-lyon1.fr) and the Sloan Digital Sky Survey IV. This research benefited from the NASA Astrophysics Data System Bibliographic Service. Additionally, we utilized the NASA/IPAC Extragalactic Database (NED), which is managed by the Jet Propulsion Laboratory at the California Institute of Technology under a contract with the National Aeronautics and Space Administration. Furthermore, we express our thanks to the ALMA Science Portal (https://almascience.nrao.edu) for its valuable resources.

**Table 3.** Shows statistics of mass dependence of the gas hydrogen fraction ($\eta$) for the sample.

| Variable | Number of valid LIRGs | Mean value | Minimum value | Maximum value | Variance | Standard deviation | Standard error |
|---|---|---|---|---|---|---|---|
| $\phi$ | 82 | 0.71 | 0.6628 | 0.7348 | 0.000137 | 0.0117 | 0.00129 |
| $\eta$ | 82 | 1.4088 | 1.3608 | 1.5087 | 0.000551 | 0.02347 | 0.00259 |
| $\log M^{*,V}$ | 82 | 10.542 | 8.715 | 12.1687 | 0.38016 | 0.61657 | 0.06809 |
| $f_{gas}$ | 82 | 0.2675 | 0.0866 | 0.8228 | 0.03 | 0.1733 | 0.01914 |

## DATA AVAILABILITY

The pipline and data in this article will be shared on reasonable request to the corresponding author.

This paper has been typeset from a T<sub>E</sub>X/L<sup>A</sup>T<sub>E</sub>X file prepared by the author.